\documentclass[apj]{emulateapj}

\shorttitle{Low Eddington Ratio Quasar at $z\sim6$}
\shortauthors{Kim et al.}

\newcommand{\ledd}{\lambda_{\rm{Edd}}}
\newcommand{\mbh}{M_{\rm{BH}}}
\newcommand{\msun}{M_{\odot}}
\newcommand{\lbol}{L_{\rm{bol}}}

\usepackage{bm}
\usepackage{amsmath}

\begin{document}

\title{The Infrared Medium-deep Survey. IV. \\
Low Eddington Ratio of A Faint Quasar at $z\sim6$: \\
Not Every Supermassive Black Hole is Growing Fast in the Early Universe}

\author{Yongjung Kim\altaffilmark{1,2}, 
Myungshin Im\altaffilmark{1,2}, 
Yiseul Jeon\altaffilmark{1,3},
Minjin Kim\altaffilmark{4,5},
Minhee Hyun\altaffilmark{1,2}, 
Dohyeong Kim\altaffilmark{1,2}, 
Jae-Woo Kim\altaffilmark{4}, 
Yoon Chan Taak\altaffilmark{1,2}, 
Yongmin Yoon\altaffilmark{1,2},
Changsu Choi\altaffilmark{1,2}, 
Jueun Hong\altaffilmark{1,2}, 
Hyunsung David Jun\altaffilmark{1,6}, 
Marios Karouzos\altaffilmark{7}, 
Duho Kim\altaffilmark{1,8}, 
Ji Hoon Kim\altaffilmark{9}, 
Seong-Kook Lee\altaffilmark{1,2}, 
Soojong Pak\altaffilmark{10},
and Won-Kee Park\altaffilmark{4}. 
}
\email{yjkim@astro.snu.ac.kr \& mim@astro.snu.ac.kr}

\affil{$^1$Center for the Exploration of the Origin of the Universe (CEOU), Building 45, Seoul National University, 1 Gwanak-ro, Gwanak-gu, Seoul 08826, Republic of Korea}
\affil{$^2$Astronomy Program, FPRD, Department of Physics \& Astronomy, Seoul National University, 1 Gwanak-ro, Gwanak-gu, Seoul 08826, Republic of Korea}
\affil{$^3$LOCOOP, Inc., 311-1, 108 Gasandigital2-ro, Geumcheon-gu, Seoul, Korea}
\affil{$^4$Korea Astronomy and Space Science Institute, Daejeon 34055, Republic of Korea}
\affil{$^5$University of Science and Technology, Daejeon 305-350, Republic of Korea}
\affil{$^6$Korea Institute for Advanced Study, 85 Hoegi-ro, Dongdaemun-gu, Seoul 02455, Korea}
\affil{$^7$Nature Astronomy, Springer Nature, 4 Crinan street, London N1 9XW, UK}
\affil{$^8$Arizona State University, School of Earth and Space Exploration, PO Box 871404, Tempe, AZ 85287-1404, U.S.A.}
\affil{$^9$Subaru Telescope, National Astronomical Observatory of Japan, 650 North A'ohoku Place, Hilo, HI 96720, U.S.A.}
\affil{$^{10}$School of Space Research, Kyung Hee University, 1732 Deogyeong-daero, Giheung-gu, Yongin-si, Gyeonggi-do 17104, Republic of Korea}


\begin{abstract}
To date, most of the luminous quasars known at $z\sim6$ have been found to be in maximal accretion with the Eddington ratios,
$\ledd\sim1$, suggesting enhanced nuclear activities in the early universe.
However, this may not be the whole picture of supermassive black hole (SMBH) growth 
since previous studies have not reached on faint quasars
that are more likely to harbor SMBHs with low $\ledd$.
To gain a better understanding on the accretion activities in quasars in the early universe,
we obtained a deep near-infrared (NIR) spectrum of a quasar, IMS J220417.92+011144.8 (hereafter IMS J2204+0112), one of the faintest quasars
that have been identified at $z\sim6$.
From the redshifted \ion{C}{4} $\lambda 1549$ emission line in the NIR spectrum, 
we find  that IMS J2204+0112 harbors a SMBH with about a billion solar mass and $\ledd \sim 0.1$,
but with a large uncertainty in both quantities (0.41 dex).
IMS J2204+0112 has one of the lowest Eddington ratios among quasars at $z\sim6$,
 but a common value among quasars at $z\sim2$.
Its low $\ledd$ can be explained with two scenarios;
the SMBH growth from a stellar mass black hole through  short-duration super-Eddington accretion events or from a massive black hole seed ($\sim10^{5}\,M_{\odot}$) with Eddington-limited accretion.
 NIR spectra of more faint quasars are needed to better understand the accretion activities of SMBHs at $z \sim 6$.

\end{abstract}


\section{INTRODUCTION}\label{sec:intro}

Since the first discovery of a quasar in 1960s, more than 400,000 quasars have been discovered by numerous surveys so far (e.g., \citealt{Schmidt83,Hewett95,Boyle00,Im07,Lee08,Shen08,Shen11,Flesch15,Paris17,Jeon17}).
Among them, about 100 quasars have been identified at $z\gtrsim6$
\citep{Fan00,Fan06,Goto06,Willott10b,Mortlock11, Banados14,Banados16,Banados17,Kashikawa15,Venemans13,Venemans15a,Venemans15b,Wu15,Kim15,Jiang09,Jiang16,Matsuoka16,Matsuoka17,Mazzucchelli17}.
Compared to quasars at lower redshifts, 
these high redshift quasars show no remarkable evolution in UV/optical spectral shapes \citep{Fan06,Jun15}, 
but a larger fraction of them is found to be dust-poor compared to their low redshift counterparts, 
a possible indication that high redshift quasars are rapidly evolving  \citep{Jiang10,Jun13}.

Using the black hole (BH) mass estimator that assumes the Doppler broadening of virialized gas as 
the dominant cause for the broad emission lines of quasars 
(e.g, see \citealt{Jun15,Kim10}),
the black hole masses ($M_{\rm{BH}}$) of few tens of high redshift quasars are found to be $10^{8-10}~M_{\odot}$ 
\citep{Jiang07,Kurk07,Kurk09,Willott10a,Mortlock11,Shen11,Jun15,Wu15,Venemans15a}. 
 Interestingly, the existence of SMBHs in such an early universe poses a theoretical challenge for the following reason.

The SMBH mass at a given time $t$ ($M_{\rm{BH}}(t)$) can be expressed as, 
\begin{equation} \label{eq:mbh}
M_{\rm{BH}}(t)= M_{\rm{BH, 0}} \times {\rm exp}\left( \dot{m} f_{\rm{Duty}}(1-\epsilon)\frac{t-t_{0}}{t_{Edd}}\right),
\end{equation}  
where $\dot{m}$ is the mass accretion rate normalized by Eddington mass accretion (see \citealt{Watarai01,Volonteri15}),
$t_{\rm{Edd}}=450$ Myr, $\epsilon$ is the radiation efficiency, $f_{\rm{Duty}}$ is the duty cycle, $M_{\rm{BH},0}$ is the seed BH mass, and $t_{0}$ is the time when the seed BH started to grow. 
For a standard disk model with Eddington-limited accretion, $\dot{m}=\ledd/\epsilon=(\lbol/L_{\rm Edd})/\epsilon$,
where $\ledd$ is the Eddington ratio, $\lbol$ is the bolometric luminosity, and $L_{\rm Edd}$ is the Eddington luminosity 
($L_{\rm{Edd}}=1.26\times10^{38}\,(M_{\rm{BH}}/M_{\odot})$ in erg s$^{-1}$).
 Adopting a typical value of $\epsilon = 0.1$ ,
even with a continuous maximal accretion at $\ledd = 1$, it requires about $\sim 0.8$ Gyr
 for a stellar mass BH with $M_{\rm{BH},0} = 100~\msun$ to grow into $10^{9} \, \msun$. 
The age of the universe is only 0.94 Gyr at $z=6$ and 0.48 Gyr at $z=10$ (a plausible redshift for a stellar mass BH to form), so the creation of a $10^{9}\,\msun$ BH is nearly impossible especially when we also 
 consider feedbacks from star formation and AGN activity that hinder the continuous Eddington-limited accretion 
  \citep{Milosavljevic09,Park12,Pelupessy07,Alvarez09,Jeon12,Johnson13}.
 To solve this problem, super-Eddington accretion ($\ledd > 1$) of stellar mass BHs (e.g., \citealt{Volonteri05,Wyithe12,Madau14}), and  BH growth from massive seed BHs with $10^{4-6}\,\msun$ are introduced (e.g., \citealt{Bromm03,Begelman06,Lodato06,Johnson13}). 

 Testing these different SMBH growth scenarios requires understanding Eddington ratios of high redshift quasars. 
 So far, the Eddington ratios are measured for about 20 luminous $z\sim6$ quasars 
 (bolometric luminosity, $\lbol \gtrsim 10^{47}$ erg s$^{-1}$) and the values are found to be at $\ledd \sim 1$ 
 (e.g., see \citealt{Willott10a,Jun15,Wu15})
 in contrast to $\ledd \sim 0.1$ of their counterparts at lower redshifts  
\citep{Richards06,Shen11,Trakhtenbrot12}.
 The predominantly Eddington-limited accretion of SMBHs at high redshift might be in line with the rapid accretion scenario in the models that allow stellar mass seed BHs 
 (e.g., see \citealt{Volonteri12,Alexander12,Johnson13} and references therein). 

 However, previous studies have been limited mostly to luminous quasars that are likely to be high $\ledd$ objects.
Therefore, the suggestion that high redshift quasars are rapidly growing could be a result of this kind of bias. To avoid the bias, \cite{Willott10a} tried to infer the intrinsic $\ledd$ distribution from  the observed $\ledd$ distribution of 17 luminous quasars at $z \sim 6$ with an assumption that the distribution follows a lognormal form. According to their analysis,
  the peak of the intrinsic $\ledd$ 
distribution of $z\sim6$ quasars is $\log(\ledd)=-0.22$, in comparison to the observed peak at 
$\log(\ledd) \sim 0.03$. This result indicates that there should be more quasars with $\ledd < 1$ 
if fainter luminosity quasars are explored, but it still implies nearly Eddington-limited accretion for 
most $z \sim 6$ quasars. 
However, recent studies of $z\sim6.5$ quasars \citep{DeRosa14,Venemans15a,Mazzucchelli17} suggested that
there are a few $10^{46.5-47}$ erg s$^{-1}$ luminous quasars with $\mbh > 10^{9.0}~\msun$, and the average $\log(\ledd)$ of 15 $z\sim6.5$ quasars is $0.39$ which is comparable to their low redshift counterparts \citep{Mazzucchelli17},
implying that the derived intrinsic $\ledd$ distribution of \cite{Willott10a} is biased toward high $\ledd$.
Also, a possible positive correlation of $\lbol$ and $\ledd$ for low redshift quasars
 \citep{Shen08,Shen11,Lusso12} may lead to the same conclusion.
Since the majority of quasars at high redshift are faint  
as implied by the quasar luminosity function (QLF; \citealt{Willott10b,Kashikawa15,Giallongo15,Kim15}),
these limited quasar sample cannot truly represent the whole quasar population at $z\sim6$, 
if $z\sim6$ quasars have such a $\lbol$-$\ledd$ relation like their low redshift counterparts.

 Thanks to the recent wide-area deep surveys, new light can be shed on the accretion activities of 
 high redshift quasars. Now, dozens of faint $z\sim6$ quasars are spectroscopically identified that have absolute magnitudes at rest-frame 1450 $\rm{\AA}$ of $M_{1450}>-24$ mag \citep{Kashikawa15,Kim15,Matsuoka16,Matsuoka17}. These faint quasars can possibly represent
  the population of low $\ledd$ SMBHs. Therefore, to really see 
  how fast high redshift quasars are growing, it is important to measure their $M_{\rm{BH}}$ and $\ledd$ values.
 So far, little has been done to characterize these faint quasars at high redshift, but deep NIR spectroscopy with
 sensitive spectrographs should be able to reveal their $\mbh$ and $\ledd$ one by one.  
 
In this paper, we present the first NIR spectroscopic observation of IMS J2204+0112
\citep{Kim15},
one of the faintest $z\sim6$ quasars discovered so far from the Infrared Medium-deep Survey
(Im et al. 2018, in preparation).  
We describe the observation and the data analysis in Section \ref{sec:obsdata}.
We present the quasar's spectral properties that are obtained through   
 continuum/line fitting in Section \ref{sec:specmodel}.
We present the $M_{\rm{BH}}$ and $\ledd$ values of IMS J2204+0112 in Section \ref{sec:results}.
The implications of our results about the growth SMBHs in the early universe are discussed in Section \ref{sec:discussion}. 
We adopt $\Omega_{m}=0.3$, $\Omega_{\Lambda}=0.7$, and $H_{0}=70$ km s$^{-1}$ Mpc$^{-1}$ of a concordance cosmology
that has been supported by observations in the past decades (e.g., \citealt{Im97}).

\section{OBSERVATION AND DATA ANALYSIS}\label{sec:obsdata}

The NIR spectroscopic observation of IMS J2204+0112 was carried out with the Folded-port InfraRed Echellette (FIRE)
mounted on the Magellan/Baade 6.5 m telescope at the Las Campanas Observatory in Chile.
The observation aimed at detecting the redshifted \ion{C}{4} line, a common $M_{\rm{BH}}$ estimator \citep{Vestergaard06,Jun15}. 
\ion{Mg}{2} is another, possibly better choice for $M_{\rm{BH}}$ measurement \citep{Shen11,Ho12,Jun15}, 
but we opted for the \ion{C}{4} line due to observational difficulty of detecting \ion{Mg}{2} at longer wavelengths.
We observed the target with the high-throughput prism mode (or longslit mode) on September 12th and 13th in 2015.
The data were taken with a $1\farcs0$ slit, which gives a spectral resolution in $J$-band ($R_{J}$) of 500,
corresponding to a resolution of $\sim 600$ km s$^{-1}$.
The single exposure time for each frame was set at 908.8 sec with the Sample-Up-The-Ramp (SUTR) readout mode,
which reads out the detector continuously during exposure.
This kind of long exposure in NIR observation makes the long wavelength region ($\lambda>12,000~\rm{\AA}$) saturated,
but enables us to obtain sufficient signals (S/N $\gtrsim3$ over a resolution element) for continuum at 
short wavelength ($\lambda<12,000~\rm{\AA}$).
We took 26 frames for IMS J2204+0112, 
but only 20 frames taken under good weather conditions (seeing $\lesssim 1\farcs0$) 
were used for the data analysis, giving a net exposure time of 5.05 hours.

\begin{figure*}
\centering
\includegraphics[width=0.8\textwidth]{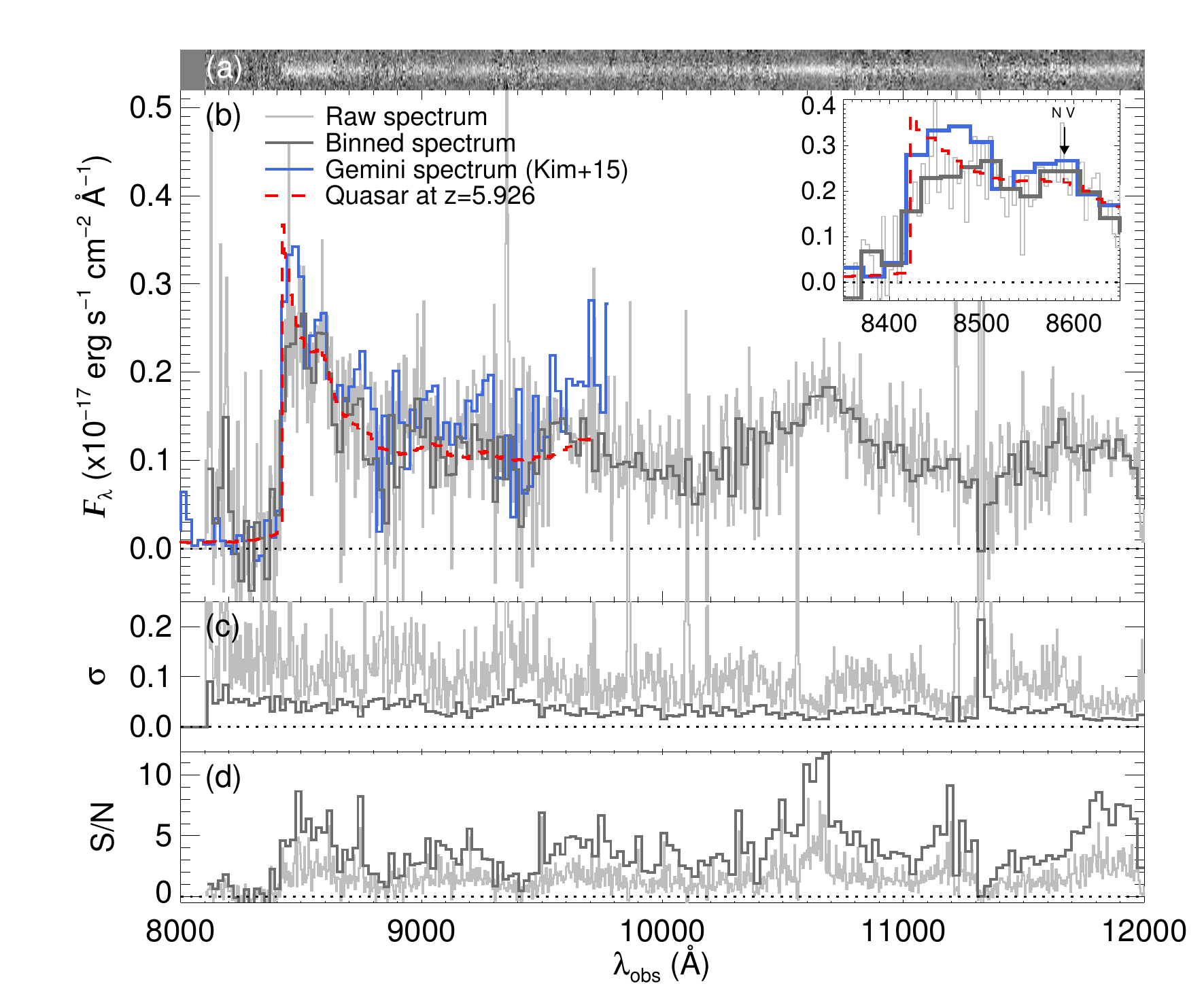}
\caption{
(a): the NIR 2D spectrum of IMS J2204+0112.
(b): the NIR spectrum of IMS J2204+0112. The light gray lines represent spectrum of IMS J2204+0112 taken with FIRE, and 
the dark gray lines show the spectrum binned at the spectral resolution of $R_{J}=500$.
The blue line represents the optical spectrum obtained with GMOS on Gemini \citep{Kim15}.
The red dashed line shows the fitted quasar model of \cite{Kim15} with $z=5.926$.
The inset shows a zoomed-in spectrum around the Lyman-$\alpha$ break at $\sim8500~\rm{\AA}$,
and we marked the peak of the \ion{N}{5} $\lambda 1240$ emission line at $z=5.926$.
(c) and (d): the spectroscopic error and S/N of the NIR spectrum, respectively.
\label{fig:obsspec}}
\end{figure*}

Although the data were obtained through nodding observation (i.e., ABBA offset),
varying seeing conditions during the observing run with long exposures generated unstable sky-lines on the spectra.
This made it difficult to eliminate the sky-lines directly by subtracting raw frames from each other.
Thus, we processed the spectra one by one, using the IRAF package \citep{Tody93}.
 Saturated regions ($\lambda>12,000~\rm{\AA}$) were trimmed, and then we performed the bias subtraction and the flat-fielding.
The wavelength solutions were derived from the NeAr arc frames.
In order to eliminate sky-lines, we subtracted median value of background pixels surrounding the target in the spatial direction
from the wavelength-calibrated, reduced spectrum, giving us clear sky-subtracted images around the target.
After combining the images, we extracted the spectrum with a $1\farcs0$ aperture.
Telluric correction with a standard star (HD 216807) was applied to the extracted 1D spectrum.
We adjusted the flux scale of the spectrum with the most recent photometric magnitude in $z$-band
from the Hyper Supreme-Cam Subaru Strategic Program (HSC SSP; \citealt{Aihara17a}), Data Release 1 \citep{Aihara17b}.
IMS J2204+0112 has $z=22.55\pm0.05$ AB mag\footnote{The $z'$-band magnitude of IMS J2204+0112 
was originally reported as 22.95$\pm$0.07 AB mag \citep{Kim15},
which is $\sim0.3$ mag fainter than the value from the HSC data, considering the difference between $z$ and $z'$ filters.
Note that this previous value is based on the images that were taken 9 years before the HSC data.
If we use this value to normalize the spectrum, it will change $\ledd$ by $\sim0.1$ dex,
which is negligible compared to other uncertainties in $\ledd$ estimate.}
in the HSC data,
giving a flux scaling factor of 0.9.
This value gives an updated $M_{1450}$ of $-23.99\pm0.05$ AB mag.
The galactic extinction was corrected by the \cite{Cardelli89} law 
with the extinction value $A_{V}$ of $\sim0.127$ \citep{Schlafly11} assuming $R_{V}=3.1$.
Figure \ref{fig:obsspec} shows the final spectrum of IMS J2204+0112.
The uncertainty of the spectrum was derived during the aperture extracting process.

\begin{figure*}
\epsscale{1.00}
\centering
\includegraphics[width=1.0\textwidth]{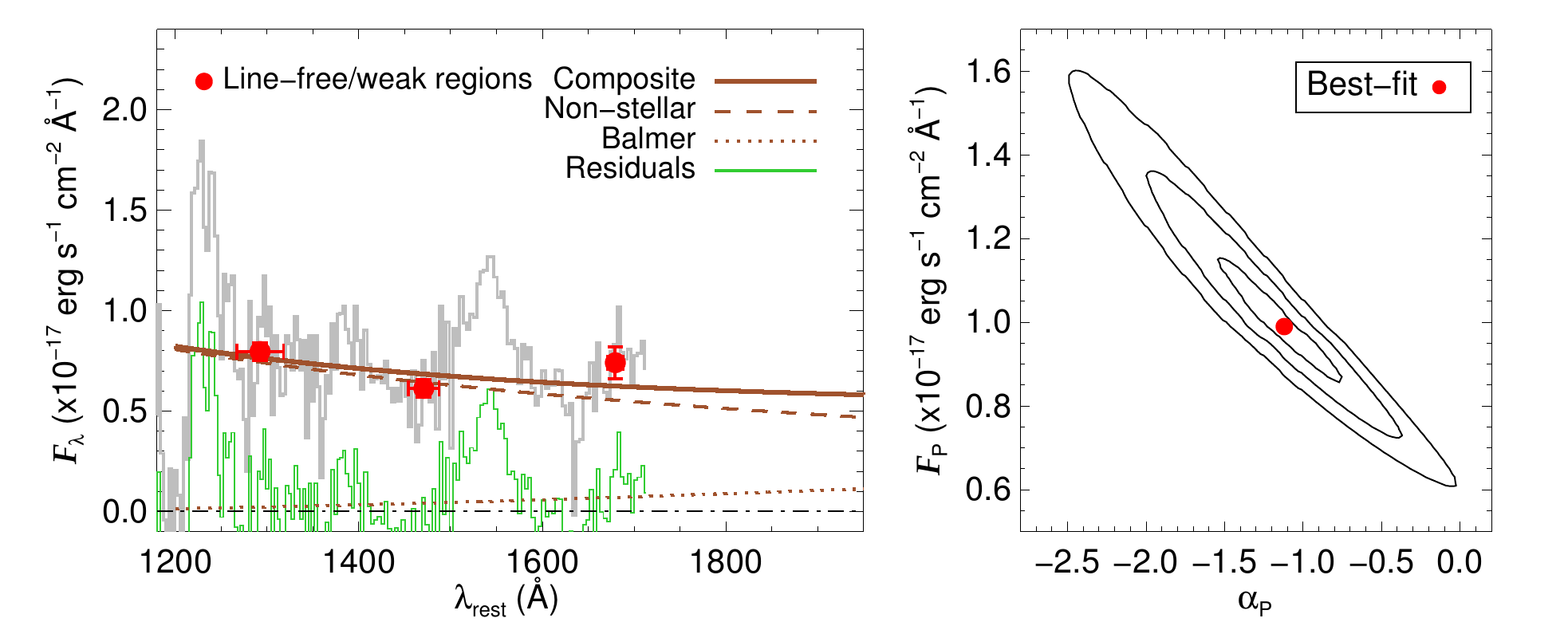}
\caption{Left: the spectrum of IMS J2204+0112 in the rest-frame.
The binned spectrum is shown as the gray line. 
The red circles represent the binned points of the spectrum at the line-free region. 
The best-fit model with the minimum $\chi^{2}_{\rm{red}}$ value is shown as the brown solid line.
This model comprises of the non-stellar power-law model (the brown dashed line) 
and the Balmer pseudo-continuum model (the brown dotted line).
The residual spectrum is shown as the green line.
Right: the parameter space of $\alpha_{\rm P}$ and $F_{\rm{P}}$ (see Section \ref{sec:continuum}).
The red dot represents our best-fit values of $\alpha_{\rm P}$ and $F_{\rm{P}}$,
and the contours show the confidence regions (1$\sigma$ to 3$\sigma$ from inner to outer).
\label{fig:cont}}
\end{figure*}

\section{SPECTRAL MODELING}\label{sec:specmodel}

In this section, we show how we performed the spectral modeling for IMS J2204+0112 to estimate its continuum 
luminosity at a specific wavelength and the full-width of half maximum (FWHM) of the \ion{C}{4} emission line.
To use a better S/N data for the spectral analysis, 
we binned the spectrum to match $R_{J}$ (the dark gray line in Figure \ref{fig:obsspec})
without overlap between the pixels used for binning.
Each bin contains 4-6 pixels, and we took weighted-mean of fluxes in each bin with
the weight of $w_{i}=\sigma_{i}^{-2}$, where $\sigma_{i}$ is the error of the $i$-th pixel in each bin.
The errors in each bin ($\sigma_{\rm bin}$) are estimated as $\sigma_{bin}=\left(\sum_{i=1}^{N_{\rm{pix}}} w_{i}\right)^{-1/2}$,
where $N_{\rm{pix}}$ is the number of pixels in each bin.
We updated the wavelength calibration of the Gemini spectrum \citep{Kim15}, and used the updated spectrum
to derive redshift, since S/N near the Lyman break is about twice larger in the Gemini spectrum than the FIRE spectrum.
Following the method described in \cite{Kim15}, we find the updated redshift value of $z=5.926\pm0.002$
by fitting a quasar spectrum model shown as red dashed line in Figure \ref{fig:obsspec}.
Note that this redshift value matches the location of the peak of \ion{N}{5} $\lambda 1240$ emission line well.

\begin{deluxetable}{cc}
\tabletypesize{\scriptsize}
\tablecaption{Continuum Fitting Results \label{tbl:cont}}
\tablewidth{0pt}
\startdata
\hline
\hline
Continuum Fitting Parameters & Best-fit Value with 1$\sigma$ error\\
\hline
$F_{\rm{P}}$ ($\times10^{-18}$ erg s$^{-1}$ cm$^{-2}$ $\rm{\AA}$$^{-1}$) & $9.90^{+1.67}_{-1.37}$ \\
$\alpha_{\rm{P}}$ & $-1.12^{+0.38}_{-0.42}$ \\
$f_{\rm{B}}$ & 1.0$^{a}$ \\
$T_{e}$ (K) &  15,000$^{b}$ \\
$\tau_{\rm{BE}}$ &  1.0$^{b}$ 
\enddata
\tablenotetext{a}{Marginal value in the fitting range. See details in Section \ref{sec:continuum}}
\tablenotetext{b}{Fixed values. See details in Section \ref{sec:continuum}}
\end{deluxetable}

\begin{figure*}
\centering
\includegraphics[width=0.8\textwidth]{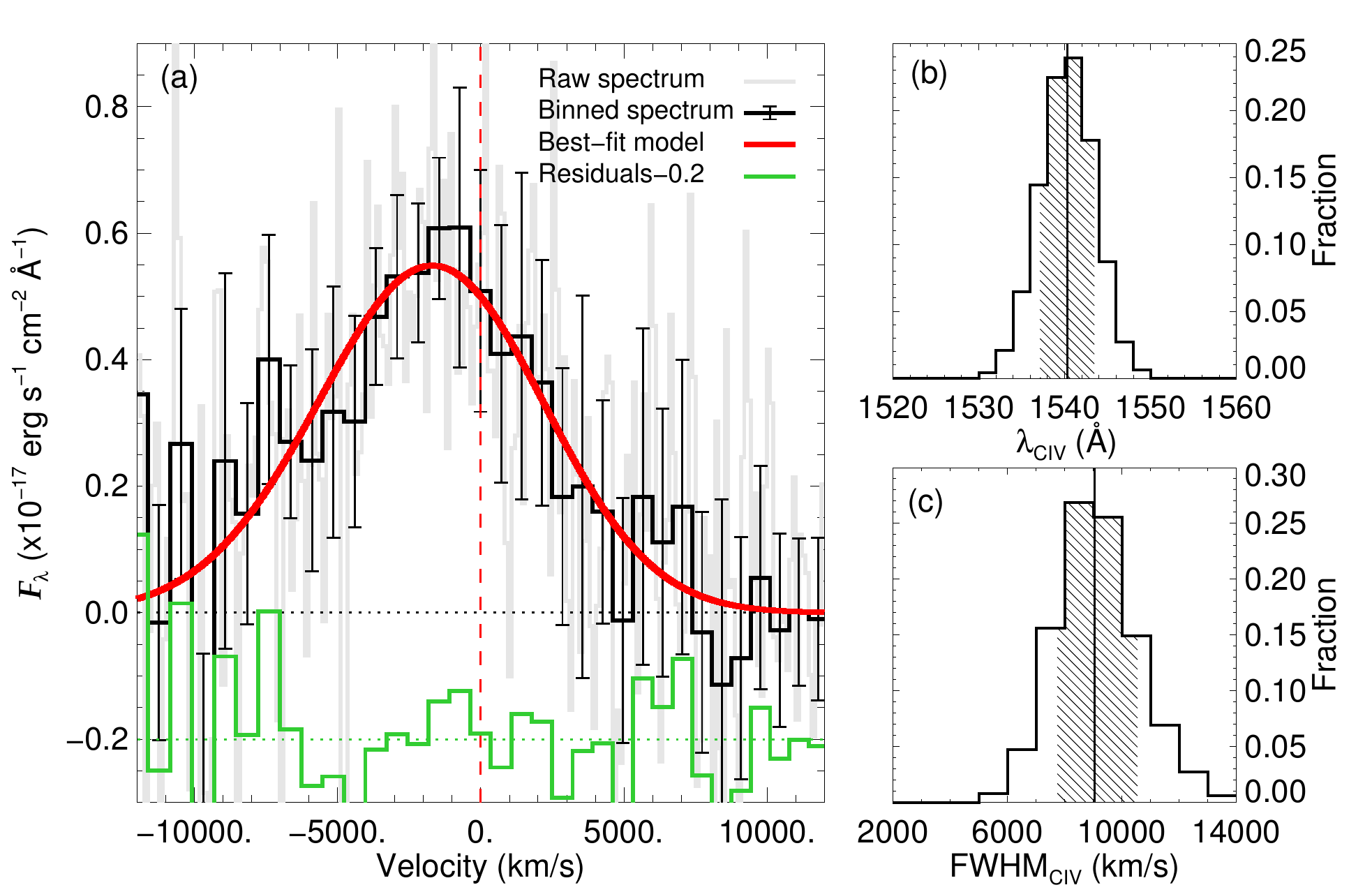}
\caption{(a) the specific flux density of \ion{C}{4} emission line of IMS J2204+0112 in rest-frame 
after subtracting the best-fit continuum model.
While the raw spectrum is shown as the gray line,
the binned spectrum with flux error is shown as the black line.
The red solid line represents the best-fit model for the \ion{C}{4} emission line,
and the green line shows the residual spectrum.
(b) and (c): The distributions of $\lambda_{\rm{CIV}}$ and FWHM$_{\rm{CIV}}$ 
in 100,000 trials, respectively.
While the vertical line in each panel indicates the best-fit result,
the shaded region corresponds to the 68\% range (or $1\sigma$ confidence level) of the distribution.
\label{fig:civ}}
\end{figure*}

\subsection{Continuum Components}\label{sec:continuum}

It is crucial for a reliable $M_{\rm{BH}}$ measurement to have a well-defined continuum model for the quasar spectrum.
To increase S/N of the continuum part of the spectrum, we binned regions with no (or weak) emission lines (e.g., 1250-1335, 1445-1495, and 1670-1690 $\rm{\AA}$) and used them (the red circles in Figure 2) to fit the continuum. 
Each binned point represents the weighted-mean value of the specific flux density in each wavelength range.
We also ignored the \ion{Fe}{2} and \ion{Fe}{3} lines in the continuum fitting, 
since they are known to be weak at $\lambda_{\rm{rest}}\lesssim2,000~\rm{\AA}$
(e.g., quasar spectra in \citealt{Jiang07,DeRosa14}). 

We modeled the quasar continuum spectrum as the sum of the non-stellar power-law continuum 
from the accretion disk and the Balmer pseudo-continuum from gas clouds surrounding the black hole as,

\begin{equation}
\begin{aligned}
F_{\lambda}=&F_{\rm{P}}\left(\frac{\lambda}{1000\rm{\AA}}\right)^{\alpha_{\rm P}} \\
 & + F_{\rm{B}}B_{\lambda}(T_{e})\left(1-e^{-\tau_{\rm{BE}}(\lambda/\lambda_{\rm{BE}})^{3}}\right),~
\lambda<\lambda_{\rm{BE}},\label{equ:cont}
\end{aligned}
\end{equation}

\noindent where $F_{\rm{P}}$ is the normalized flux density for the non-stellar power-law continuum at 1000 $\rm{\AA}$,
$\alpha_{\rm P}$ is the slope of the power-law continuum, $F_{\rm{B}}$ is the normalized flux density for the Balmer continuum,
$B_{\lambda}(T_{e})$ is the Planck function at an electron temperature $T_{e}$, 
and $\tau_{\rm{BE}}$ is the optical depth at the Balmer edge ($\lambda_{\rm{BE}}=3646~\rm{\AA}$ in the rest frame;
\citealt{Grandi82}).
Since both high and low redshift quasars have the slope of $\alpha_{\rm P}=-1.5\pm1.2$
\citep{DeRosa11,Decarli10,Shen11}, we adopted the fitting range of $-3.0 \le \alpha_{\rm P} \le 1.0$,
which covers 1$\sigma$ dispersion of $\alpha_{\rm P}$.
The second term is for the Balmer pseudo-continuum from \cite{Dietrich03}.
The basic assumption is that there are gas clouds with uniform $T_{e}=15,000$ K \citep{Dietrich03}
in a partially optically thick condition ($\tau_{\rm{BE}}=1.0$; \citealt{Kurk07}).
We also tested models with $10,000 \le T_{e} \le 20,000$ K and $0.1 \le \tau_{\rm{BE}} \le 2.0$,
the range that previous studies used (e.g., \citealt{DeRosa14}), 
but there are no significant differences between the models
due to the small contribution of the Balmer continuum to the composite continuum at $\lambda_{rest}<2000~\rm{\AA}$.
Since our NIR spectrum does not cover the wavelength ($\lambda_{rest}=3675~\rm{\AA}$)
where the normalization of the model is usually done 
\citep{Dietrich03,Kurk07,Jiang09,DeRosa11,DeRosa14}, 
we normalized the Balmer continuum with assumptions that 
(i) the power-law continuum is dominant at our fitting range of 
$1200<\lambda_{rest}<1800~\rm{\AA}$, and 
(ii) the flux density of the Balmer continuum can be normalized to a fraction of the
power-law continuum flux density at $\lambda_{rest}=3675~\rm{\AA}$ that is extrapolated from our NIR data: 
$F_{\rm{B}}=f_{\rm{B}}\cdot F_{\rm{P}}\cdot(3675~\rm{\AA})^{\alpha_{\rm P}}$,
where $f_{\rm{B}}$ is the fraction of the Balmer continuum at $3675~\rm{\AA}$.
Since $f_{\rm{B}}$ is less than 1.0 and typically $\sim0.3$
\citep{Dietrich03,DeRosa11},
the fitting range of $f_{\rm{B}}$ is set to $0.1 \le f_{B} \le 1.0$.

We calculated $\chi_{\rm{red}}^{2}$ values with a
grid-based parameter set of ($F_{\rm{P}}$, $\alpha_{\rm P}$, $f_{\rm{B}}$),
and found the best-fit result that has the minimum $\chi_{\rm{red}}^{2}$ value,
given in Table \ref{tbl:cont}.
The errors were computed by finding marginal points of $\chi_{\rm{red}}^{2}<\chi_{\rm{red,min}}^{2}+1$ 
(1$\sigma$ confidence level) in the parameter space.
Figure \ref{fig:cont} shows the best-fit continuum model plotted on the NIR spectrum of IMS J2204+0112.
The best-fit non-stellar power-law model has a slope of
$\alpha_{\rm P}=-1.12^{+0.38}_{-0.40}$, consistent with that of other high redshift quasars.
For the Balmer pseudo-continuum model,
the best-fit model results in $f_{\rm{B}}=1.0$ due to the significant flux at $\sim1680~\rm\AA$.

The flux density of the best-fit continuum model and its 1$\sigma$ error are generated from
$\chi^{2}$ distribution of $\alpha$ and $F_{\rm{P}}$ (Figure \ref{fig:cont}), 
while the other parameters ($f_{\rm{B}}$, $T_{e}$, and $\tau_{\rm{BE}}$) are fixed.
From the flux density of the best-fit continuum model in the rest-frame system,
we calculated the monochromatic continuum luminosity at $\lambda_{rest}=1350~\rm{\AA}$ and $1450~\rm{\AA}$
($L_{1350}$ and $L_{1450}$, respectively), assuming isotropic radiation at the luminosity distance of IMS J2204+0112.
We also computed the bolometric luminosity $\lbol$ from $L_{1450}$,
using the quasar bolometric correction from \cite{Runnoe12}: $\lbol=4.20 \times L_{1450}$.
The estimated values with the errors in $1\sigma$ confidence level are given in Table \ref{tbl:phys}.
The $\log(\lbol$) of IMS J2204+0112 is only $46.21^{+0.12}_{-0.16}$ erg s$^{-1}$.
Note that the errors from both the flux density and the best-fit continuum model are included in the uncertainty.

\begin{deluxetable}{cc}
\tabletypesize{\scriptsize}
\tablecaption{Spectral Properties of IMS J2204+0112 \label{tbl:phys}}
\tablewidth{0pt}
\startdata
\hline
\hline
Estimated Properties & Best-fit Value with 1$\sigma$ error\\
\hline
$z$$^{a}$ & 5.926$\pm$0.002 \\
$\log L_{1350}$ (erg s$^{-1}$) & $45.59^{+0.08}_{-0.10}$ \\
$\log L_{1450}$ (erg s$^{-1}$) & $45.59^{+0.12}_{-0.16}$\\
$\log \lbol$ (erg s$^{-1}$) & $46.21^{+0.12}_{-0.16}$\\
$\lambda_{\rm{CIV}}$ ($\rm{\AA}$) & $1540.32^{+3.14}_{-3.20}$\\
FWHM$_{\rm{CIV}}$ (km s$^{-1}$) & $9046^{+1499}_{-1305}$\\
$\sigma_{\rm{G}}$ (km s$^{-1}$) & $3841^{+636}_{-554}$
\enddata
\tablecomments{The uncertainties of luminosity are lower limits with constraining the contribution of
the Balmer pseudo-continuum and elimination of iron lines for fitting.}
\tablenotetext{a}{Derived from Gemini spectrum \citep{Kim15}}
\end{deluxetable}

\begin{deluxetable*}{lcccc}
\tabletypesize{\scriptsize}
\tablecaption{$M_{\rm{BH}}$ and $\ledd$ of IMS J2204+0112 \label{tbl:mbh}}
\tablewidth{500pt}
\startdata
\hline
\hline
Reference & Method & $\log(M_{\rm{BH,CIV}}/M_{\odot})$ & $\sigma_{int}$ & $\log \ledd$ \\
(1) & (2) & (3) & (4) & (5) \\
\hline
\cite{Vestergaard06}$^{\dagger}$ & $\gamma=2$ & $9.38^{+0.13}_{-0.15}$ & 0.36 & $-$1.27 \\ 
\cite{Jun15} & $\gamma=2$ & $9.55^{+0.24}_{-0.24}$ & 0.40 & $-$1.43 \\ 
\cite{Park17} & $\gamma=2$ & $9.27^{+0.19}_{-0.20}$ & 0.22 & $-$1.16 \\ 
\cite{Park17}$^{\dagger}$ & $\gamma=0.50$ & $8.72^{+0.60}_{-0.59}$ & 0.16 & $-$0.61 \\ 
\cite{Coatman17}$^{\dagger}$ & $v_{bs,{\rm{CIV}}}$$^{\ddagger}$ & $9.05^{+0.26}_{-0.29}$ & $\sim$0.5 & $-$0.93 \\ 
\cite{Jun17} & $v_{bs,{\rm{CIV}}}$$^{\ddagger}$ & $9.27^{+0.27}_{-0.28}$ & $\sim$0.35 & $-$1.15 \\ 
\cite{Park17}$^{\dagger}$ & $\sigma_{\rm{CIV}}$ & $8.59^{+0.19}_{-0.21}$ & 0.12 & $-$0.48 \\ 
\cite{Park17} & $\sigma_{\rm{G}}$ & $8.58^{+0.18}_{-0.19}$ & 0.12 & $-$0.47 \\ 
\hline
Weighted-mean & - & $9.09\pm0.41$ & - & $-$0.97  
\enddata
\tablecomments{
The results of $M_{\rm{BH,CIV}}$ and $\ledd$ measurements by several methods.
Column 1: References
Column 2: Methods for $M_{\rm{BH,CIV}}$ estimation.
Column 3:  $M_{\rm{BH,CIV}}$ with $1\sigma$ errors. 
The intrinsic scatter of each method is not included in the error.
Column 4: Intrinsic scatter of $M_{\rm{BH}}$ estimator.
Column 5: $\ledd$.
}
\tablenotetext{$\dagger$}{The methods used for calculating the weighted-mean $M_{\rm{BH}}$ value with the weight of the inverse variance of the $M_{\rm{BH}}$ estimates.}
\tablenotetext{$\ddagger$}{The $v_{bs,\rm CIV}$ value used in this method is derived from the continuum break and the \ion{N}{5} line, and this procedure could bias the result.}
\end{deluxetable*}

\subsection{\ion{C}{4} Line Measurement\label{sec:line}}

After subtracting the best-fit continuum model obtained from Section \ref{sec:continuum},
we fitted the \ion{C}{4} emission line and measured its spectral properties.
It is well-known that the \ion{C}{4} emission line of quasars often shows asymmetric line shapes 
that cannot be well modeled by a single Gaussian function 
\citep{Shen11,Tang12,Runnoe13,Park13,Park17,DeRosa14,Karouzos15,Coatman16}.
While this asymmetric line shape of \ion{C}{4} can be seen in high S/N spectra 
(S/N $\gtrsim10$ for continuum), 
it is not discernible in the spectrum with low S/N of $\lesssim10$ \citep{DeRosa14} like our case.
Hence, the \ion{C}{4} emission of IMS J2204+0112 was fitted with a single Gaussian function.
For the error analysis, we adjusted the parameters of the non-stellar power-law continuum ($F_{\rm{P}}$ and $\alpha_{\rm P}$)
by using random pairs of $\alpha_{\rm P}$ and $F_{\rm{P}}$ following the $\chi^{2}$ distribution in parameter space
(the right panel in Figure \ref{fig:cont}).
This process enables us to determine error of the continuum flux density per binned pixel.  
We took the quadratic sum of errors of the continuum model and of the NIR spectrum
as the uncertainties of the continuum-subtracted spectrum for each pixel.

We used the MPFIT package \citep{Markwardt09},
a robust non-linear least squares curve fitting 
with the Levenberg-Marquardt technique, for the \ion{C}{4} line fitting.
The fitting range was set to
$1400~\rm{\AA} \le \lambda_{rest} \le 1650~\rm{\AA}$.
The fitting provides the central peak wavelength $\lambda_{\rm{CIV}}$, and the Gaussian standard deviation 
$\sigma_{\rm{G}}$ that is converted to the \ion{C}{4} line FWHM (FWHM$_{\rm{CIV}}$) with a relation of
FWHM$\simeq2.355\times \sigma_{\rm{G}}$.
Note that the instrumental resolution of FWHM$_{ins}=600$ km s$^{-1}$ is subtracted from the measured FWHM$_{obs}$ as 
FWHM$_{\rm{CIV}}=\sqrt{({\rm{FWHM}}_{obs})^{2}-({\rm{FWHM}}_{ins})^{2}}$.

The panel (a) in Figure \ref{fig:civ} shows the radial velocity profile of the \ion{C}{4} line.
The red solid line indicates the best-fit model for the \ion{C}{4} emission line with 
$\lambda_{\rm{CIV}}=1540.32^{+3.14}_{-3.20}~\rm{\AA}$
and FWHM$_{\rm{CIV}}=9046^{+1499}_{-1305}$ km s$^{-1}$ (or $\sigma_{\rm{G}}=3841^{+636}_{-554}$ km s$^{-1}$).
To derive the errors, 
we generated 100,000 mock radial profiles by adding appropriate random Gaussian noises to the best-fit model.
After re-fitting the mock spectra,
we took the 68\% ranges of the distributions of $\lambda_{\rm{CIV}}$ and FWHM$_{\rm{CIV}}$ as their 1$\sigma$ errors 
(panels (b) and (c) in Figure  \ref{fig:civ}).

\begin{figure*}
\centering
\includegraphics[width=1.0\textwidth]{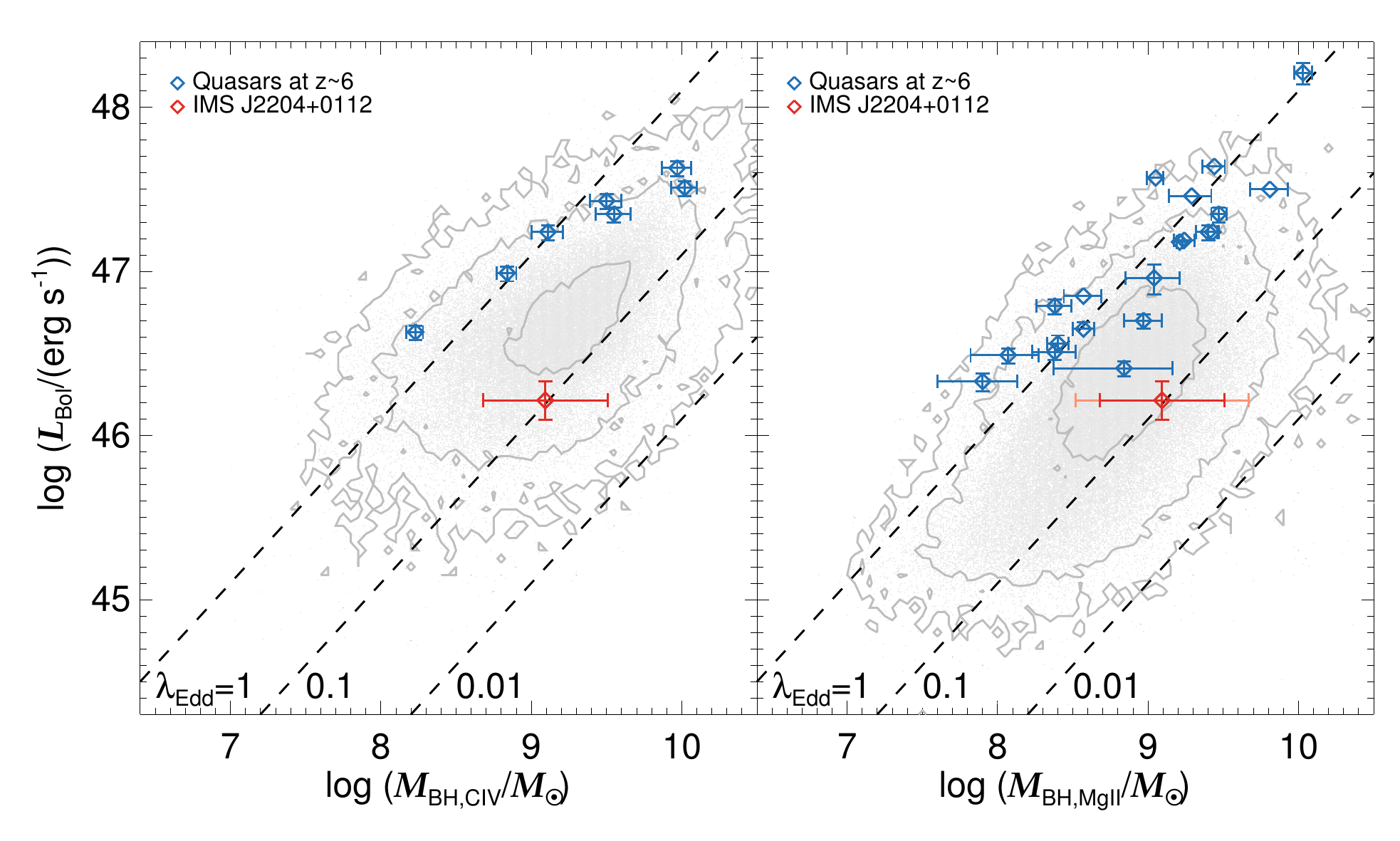}
\caption{
The $M_{\rm{BH}}$-$\lbol$ distributions of quasars.
The left and the right panels show the results based on $M_{\rm{BH,CIV}}$ and $M_{\rm{BH,MgII}}$, respectively.
While the gray dots and the contours represent the low redshift quasars from SDSS DR7 Quasar catalog \citep{Shen11},
the blue diamonds indicate quasars at $z\sim6$ \citep{Jiang07,Kurk07,Kurk09,Willott10a,DeRosa11,Wu15}.
IMS J2204+0112 is shown as the red diamond,
which seems to be isolated from other high redshift quasars.
Note that the error bar of IMS J2204+0112 with light red color in the right panel includes the error of $\mbh$
measurements and the dispersion of $M_{\rm{BH,MgII}}$ compared to $M_{\rm{BH,CIV}}$.
This figure indicates that IMS J2204+0112 is a quasar with an exceptionally low $\ledd$ among $z=6$ quasars.
\label{fig:mbhlbol}}
\end{figure*}

\section{RESULTS}\label{sec:results}

\subsection{Black Hole Mass\label{sec:mbh}}

The BH mass, $M_{\rm{BH,CIV}}$ of IMS J2204+0112 is estimated using scaling relations that utilize $L_{1350}$ and FWHM$_{\rm{CIV}}$ as below: 

\begin{equation}
\begin{aligned}
{\log} & \left( \frac{\textit{M}_{\rm{BH,CIV}}}{\textit{M}_{\odot}} \right)=  \\
&A   +{\log} \left\lbrace \left(\frac{L_{1350}}{10^{44}~\rm{erg~s^{-1}}} \right)^{\beta} 
\left( \frac{\rm{FWHM_{CIV}}}{1000~\rm{km~s^{-1}}} \right)^{\gamma} \right\rbrace. 
\label{equ:mbh}
\end{aligned}
\end{equation}

 Many groups have suggested that one needs to be cautious about $M_{\rm{BH,CIV}}$. 
 The $M_{\rm{BH,CIV}}$ values are found to have a large scatter of $\sim0.4$ dex against H$\beta$ or \ion{Mg}{2} based $M_{\rm{BH}}$ values \citep{Vestergaard06,Shen11,Ho12,Jun15,Jun17}.
 Also, the \ion{C}{4} line often shows an asymmetric shape possibly due to non-virial motion of gas and/or blending with
other neighboring lines, suggesting that
virial motions may not be the dominant component that determines the \ion{C}{4} line width. The unusual line shape is often associated with the blueshift of the \ion{C}{4} line which is thought to be one of the main uncertainties in the \ion{C}{4}-based estimator. Several new $M_{\rm{BH}}$ estimators are derived to use blueshift as a way to improve $\mbh$ measurements  \citep{Coatman16,Jun17}. 
Considering these various ways of obtaining $M_{\rm{BH}}$ from the \ion{C}{4} line, we derived $M_{\rm{BH, CIV}}$ using several representative estimators.
Note that the virial factor of $\log f=0.71$ \citep{Woo13} was used.
  
First, we used the estimators consistent with the idea that the exponent of the velocity term reflects the virial motion of the broad line region gas, i.e., $\gamma \sim 2$. For this, we adopted the $M_{\rm{BH, CIV}}$ estimator of \citet{Vestergaard06}, \citet{Jun15}, and \citet{Park17} where the parameter set values $(A, \beta, \gamma)$ are (6.66, 0.53, 2.0), (6.707, 0.547, 2.11), and (6.84, 0.33, 2.00) respectively. The intrinsic scatters in the derived $M_{\rm{BH}}$ are 
of order of $\pm 0.3$ dex in these estimators (see Table \ref{tbl:mbh}). 
Using the line luminosity and FWHM values we obtained in Section \ref{sec:specmodel},
we find that the $ M_{\rm{BH,CIV}}$ values of IMS J2204+0112 are 
$\log(M_{\rm{BH,CIV}}/M_{\odot})=9.38^{+0.13}_{-0.15}$ \citep{Vestergaard06}, $9.55^{+0.24}_{-0.24}$ \citep{Jun15}, 
 and $9.27^{+0.19}_{-0.20}$ \citep{Park17}. 
The $1\sigma$ uncertainty of $M_{\rm{BH,CIV}}$ is estimated by inserting the rms uncertainties
of $L_{1350}$ and FWHM$_{\rm{CIV}}$ in the $M_{\rm{BH}}$ estimators.
 All the three estimators give values that are consistent within error, with $\log(M_{\rm{BH,CIV}}/M_{\odot}) \sim 9.4$.

Second, we used the estimator with a very small $\gamma$ value of $\sim 0.5$ which is not consistent with the virial motion assumption. This kind of estimator is put forward to minimize the scatter in $\mbh$ between this method and the reverberation mapping result. Using the relation that adopts a parameter set of (7.54, 0.45, 0.5) from \citet{Park17}, we find 
$\log(M_{\rm{BH,CIV}}/M_{\odot})=8.72^{+0.60}_{-0.59}$ with an intrinsic scatter of 0.16 dex.
This is about 0.6 dex smaller than the nominal $M_{\rm{BH}}$ estimates above,
but showing very large uncertainty due to $\gamma$ of $0.50^{+0.55}_{-0.53}$. 
However, the adoption of the low $\gamma$ value may not be physically plausible \citep{Denney13}, 
and \citet{Jun15} have shown that such a relation is likely to underestimate/overestimate $M_{\rm{BH}}$ 
at high $(\log(M_{\rm{BH}}/M_{\odot})>9.5$) and low mass end $(\log(M_{\rm{BH}}/M_{\odot})<8$).

Third, we used the estimators that correct the blueshift effect of the \ion{C}{4} line, 
since the blueshift of \ion{C}{4} line ($v_{bs,\rm{CIV}} \equiv c\times(1549.48-\lambda_{\rm{CIV}})/1549.48$)
can be an indicator to correct possible bias in $M_{\rm{BH,CIV}}$ \citep{Coatman16,Coatman17,Jun17}.
 Using the $\lambda_{\rm{CIV}}$ value from Section \ref{sec:line} and the systemic redshift of $z=5.926$,
we estimate the \ion{C}{4} blueshift as $v_{bs,{\rm{CIV}}}=1685^{+608}_{-620}$ km s$^{-1}$.
Using either the parameter set of (6.71, 0.53, 2) in Eq. (6) of \citet{Coatman17} or  
$M_{\rm{BH, CIV}}$ with the blueshift correction term of \citet{Jun17}, 
we get $\log(M_{\rm{BH,CIV}}/M_{\odot})=9.05^{+0.26}_{-0.29}$, and $\log(M_{\rm{BH,CIV}}/M_{\odot})=9.27^{+0.27}_{-0.28}$, respectively. These values are consistent within error.
Note that the systemic redshift of IMS J2204+0112 is derived from the continuum break and the location of the \ion{N}{5} line, we assume that this is identical to the redshift derived from a narrow high ionization line (e.g., [\ion{O}{3}]),
or host galaxy emission (e.g., Far-infrared [\ion{C}{2}]). If this assumption is wrong, the derived $\mbh$ with this method could be biased.
Furthermore, the \ion{Mg}{2} line of a few high redshift quasars is statistically blueshifted compared to CO and [\ion{C}{2}]
emission lines, while that of low redshift ones is not \citep{Venemans16,Mazzucchelli17}.
These imply that the application of the blueshift correction factor from the $z<4$ quasars
may be inappropriate for high redshift quasars.

An alternative way to derive $M_{\rm{BH,CIV}}$ is to use line dispersion of \ion{C}{4} line ($\sigma_{\rm{CIV}}$;
\citealt{Denney13,Park13,Park17}).
The second moment line dispersion $\sigma_{\rm{CIV}}$
is $\sim3900\pm700$ km s$^{-1}$ which is calculated within $\pm10000$ km s$^{-1}$ around $\lambda_{\rm{CIV}}$.
With the best-fit parameter set from \cite{Park17},
this $\sigma_{\rm{CIV}}$ and the $\sigma_{\rm{G}}$ (estimated in Section \ref{sec:line})
give $\log(M_{\rm{BH,CIV}}/M_{\odot})=8.59^{+0.19}_{-0.21}$ and $8.58^{+0.18}_{-0.19}$, respectively.
But the $\sigma_{\rm{CIV}}$ value varies significantly with the fitting range due to the low S/N of continuum,
as also noticed in previous studies \citep{Denney13,Coatman16}.
Furthermore, $M_{\rm{BH,CIV}}$ with $\sigma_{\rm{G}}$ is possibly underestimated 
considering a common shape of \ion{C}{4} line \citep{Denney13,Park13,Park17}.

In Table \ref{tbl:mbh}, we list these $M_{\rm{BH, CIV}}$ values of IMS 2204+0112.
As a representative $\mbh$ value, we use the weighted-mean of $\mbh$ value
$(\log (M_{\rm{BH,CIV}}/M_{\odot})=9.09\pm0.41$)
from different methods;
$\gamma=2$ \citep{Vestergaard06}, $\gamma=0.5$ \citep{Park17}, $v_{bs,\rm{CIV}}$ \citep{Coatman17}, 
and $\sigma_{\rm{CIV}}$ \citep{Park17}.
Note that the weight is the inverse-variance of $\mbh$ estimation in each method.
Not surprisingly, this value matches closely with the $\mbh$ value from \ion{Mg}{2} of lower redshift quasars 
with spectral characteristics similar to IMS J2204+0112\footnote{
One can also adopt $\mbh$ derived from \ion{Mg}{2} estimators of quasars that have spectral properties
similar to IMS J2204+0112. For this, we selected quasars with $7500 < \rm{FWHM}_{\rm{CIV}}~(\rm{km~s^{-1}}) < 10500$ and
$45<\log L_{1350}~(\rm{erg~s^{-1}})<46$ from \cite{Shen11} and obtained their mean $\mbh$ from \ion{Mg}{2}.
We obtain $\log (M_{\rm{BH,MgII}} / M_{\odot})=9.08\pm0.40$.
}.

\subsection{Eddington Ratio\label{sec:intrinsic}}

Using the $M_{\rm{BH,CIV}}$ and $\lbol$ values from previous sections, 
we calculate $\ledd=\lbol/L_{\rm{Edd}}$.
The calculated $\ledd$ values are listed in Table \ref{tbl:mbh}, indicating that $\ledd$ is $0.10$,
one of the lowest values among quasars at $z \sim 6$. 

 Figure \ref{fig:mbhlbol} shows $\lbol$ versus $M_{\rm{BH}}$ of IMS J2204+0112 (the red diamond; weighted-mean $M_{\rm{BH,CIV}}$ value), 
quasars at $z \sim 6$ (the navy diamonds) and at $z < 3$ (the gray dots and contours). 
 On the left panel, we show the values that are based on $M_{\rm{BH, CIV}}$ from the \citet{Vestergaard06} relation, 
and on the right panel, the \ion{Mg}{2}-based $\mbh$ values, $M_{\rm{BH,MgII}}$ \citep{Vestergaard09}, are given.
 The $\lbol$ and $\mbh$ values of  $z\sim6$ quasars are derived in the same manner as IMS J2204+0112 
using the literature values of $L_{1350}$ and FWHM$_{\rm{CIV}}$ \citep{Jiang07,Kurk07} 
or $L_{3000}$ and FWHM$_{\rm{MgII}}$ \citep{Willott03,Willott10a,Kurk07,Kurk09,DeRosa11,Wu15}. 
For quasars at $z<3$, we take the values from \citet{Shen11} where 
the $M_{\rm{BH,CIV}}$ values are based on the \citet{Vestergaard06} relation and the $M_{\rm{BH,MgII}}$ values are derived using the \citet{Vestergaard09} relation.  

 The striking feature in the figure is that IMS J2204+0112 occupies a unique parameter space,  the parameter space 
 that has not been populated by other $z=6$ luminous quasars, but a rather common among $z \sim 2$ quasars. 
 This prompts a question as if we have been seeing only a limited population of 
 high  $\ledd$ quasars  in previous studies. 
If we impose the survey depth of IMS of $J_{\rm{AB}}<22.5-23.0$ mag \citep{Kim15}
for the intrinsic $\ledd$ distribution from \cite{Willott10a},
the $\ledd$ distribution for such a magnitude-limited survey 
would have a peak value at $\log \ledd=-0.10$ and the dispersion of 0.26 dex.
In such a case, there is only a chance of $\sim0.03\%$ (or 3.5$\sigma$ away from the peak)
to find a quasar with $\ledd$ lower than IMS J2204+0112. 
Even if we consider 1$\sigma$ error of $\ledd$ of IMS J2204+0112 ($\log \ledd = -0.56$),
the probability is only 3.84\% which is still low.
That is to say, the probability of finding such a quasar in IMS is quite low for the intrinsic $\ledd$ distribution of \cite{Willott10a}.

\section{DISCUSSION  }\label{sec:discussion}

It is remarkable that there is a faint quasar with only $\ledd = 0.10$ at $z\sim6$,
though its mass determination is quite uncertain due to the characteristic of \ion{C}{4} line.
Recently, it has been suggested that the average $\ledd$ of high redshift quasars is similar to that of their luminosity-matched counterparts at low redshift \citep{Mazzucchelli17}.
The existence of IMS J2204+0112 reinforces the recent suggestion
even at a lower $\lbol$ of $\sim 10^{46}$ erg s$^{-1}$.

As we mentioned in the introduction, the growth of 100 $M_{\odot}$ seed BH to a $\sim10^{9} M_{\odot}$ SMBH at $z=6$
is already very challenging due to the short time available between the creation of the BH seed and the epoch of $z=6$.
The situation gets significantly worse if $\ledd = 0.10$.
At $\ledd = 0.10$, Eq. (\ref{eq:mbh}) shows that it takes 8 Gyr to obtain a $10^{9}\,M_{\odot}$ BH from a stellar mass seed. Therefore, in such a case, it is impossible to grow stellar mass BHs into SMBHs in quasars at $z \sim 6$. Thus, alternative scenarios must be sought for if $\ledd$ value is around 0.10 for IMS J2204+0112 at $z\sim6$.

 Recent studies promote super-Eddington accretion as a way to create $10^{9}~M_{\odot}$ BHs by $z=6$.
 In that scenario, episodes of short duration or steady super-Eddington accretion are shown to create SMBHs by $z=6$, with a duty-cycle of 0.5 or less
 \citep{Smole15,Volonteri15,Madau14,Pezzulli16,Li12,Sakurai16,DeGraf17}. 
In such a case for the super-Eddington accretion with a slim disk \citep{Watarai01,Wang03,Ohsuga05,Volonteri15}, $\dot{m}$ in Eq. (\ref{eq:mbh}) is given by
\begin{equation}
\dot{m}\sim \frac{2}{\epsilon} \exp \left( \frac{\ledd}{2} -1 \right),\label{equ:super}
\end{equation}
for $\ledd \geq 2$.
For example, if we have a super-Eddington accretion with $\ledd=3$, adopting $\epsilon \sim 0.04$ \citep{Mineshige00} with
a duty cycle of $f_{\rm{Duty}}=0.5$, only about $180$ Myr is needed to create a $10^{9}~M_{\odot}$ BH from a $10^{2}~\msun$ seed BH, while the SMBH can have a low $\ledd$ ($\sim 0.1$ or less) in the other half of time.
Under the episodic super-Eddington accretion scenario with a stellar mass seed BH, our result of $\ledd = 0.1$ implies that IMS J2204+0112 underwent bursts of super-Eddington accretion before, and be relatively quiescent at $z\sim6$.

  Another possible BH growth scenario is to have very heavy seed BHs with $M_{\rm{BH},0} \sim 10^{4}$ to $10^{6}~M_{\odot}$
   (\citealt{Volonteri08,Johnson13,Smidt17,DeGraf12,DiMatteo12,Johnson13,Ferrara14,Gallerani17,Regan17,
   Pacucci15,Gallerani17,Regan17} and references therein).
   Using Eq. (\ref{eq:mbh}) with the final BH mass of $M_{\rm{BH}} = 10^{9} M_{\odot}$, 
   and $M_{\rm{BH},0} = 10^{5}~M_{\rm{BH}}$ for the seed BH, we get the accretion time scale of $\sim4.6$ Gyr if the accretion continues at $\ledd=0.10$ and $\sim 0.46$ Gyr at $\ledd = 1$.
   Therefore, a $10^{5}\,\msun$ seed BH can become a $10^{9}\,\msun$ BH if the BH growth can last about a few hundred Myr at the Eddington limit before subsiding to $\ledd \sim 0.1$ at $z=6$. 
  Simulations show that cold gas flows can feed massive BH seeds \citep{Smidt17,DeGraf12,DiMatteo12}. In the simulation, the BH growth proceeds nearly at Eddington-limited accretion for an extended period until $z \sim 7$ or so and then reduces to $\ledd \sim 0.1$ or less (e.g, \citealt{DiMatteo12,Smidt17}). This is consistent with our finding.

\section{Conclusion}

Through deep NIR spectroscopic observation using FIRE on the Magellan telescope, 
 we measured $\mbh$ and $\ledd$ of one of the faintest quasars at $z \sim 6$. Our result shows that IMS J2204+0112 has 
 $\mbh \sim 10^{9} \msun$ and a relatively low Eddington ratio of $\ledd = 0.1$ in comparison to other $z=6$ quasars, implying that IMS J2204+0112 is a mature SMBH at high redshift with two possible growth scenarios; the BH growth from  massive seed BH ($\sim10^{5}\,\msun$), or the BH growth through short, episodic super-Eddington accretion of stellar mass BHs.
The rather low $\ledd$ of IMS J2204+0112 is in line with the recent report that the average $\ledd$ of high redshift quasars could be similar to that of lower redshift quasars \citep{Mazzucchelli17}.
The reliability of the $\mbh$ measurements can be improved by observing the \ion{Mg}{2} line or the Balmer lines, 
 and the $\ledd$ measurements can be improved with multi-wavelength observation that includes longer wavelengths (e.g., submm). The upcoming extremely large telescopes, such as the Giant Magellan Telescope and the James-Webb Space Telescope, will allow us to routinely observe faint quasars to measure $M_{\rm{BH}}$ reliably, giving a vivid perspective for the SMBH evolution in the early universe.

\acknowledgments

We would like to thank K. Ohsuga and J. -M. Wang for useful discussions.
This work was supported by the National Research Foundation of Korea (NRF) grant, 
No. 2017R1A3A3001362, funded by the Korea government.
This work was supported by K-GMT Science Program (PID:gemini\_KR-2015A-023) of Korea Astronomy and Space Science Institute (KASI).
Based on observations obtained at the Gemini Observatory acquired through the Gemini Science Archive and processed using the Gemini IRAF package, which is operated by the Association of Universities for Research in Astronomy, Inc., under a cooperative agreement with the NSF on behalf of the Gemini partnership: the National Science Foundation 
(United States), the National Research Council (Canada), CONICYT (Chile), the Australian Research Council (Australia), Minist\'{e}rio da Ci\^{e}ncia, Tecnologia e Inova\c{c}\~{a}o (Brazil) and Ministerio de Ciencia, Tecnolog\'{i}a e Innovaci\'{o}n Productiva (Argentina).
This paper includes data gathered with the 6.5 meter Magellan Telescopes located at Las Campanas Observatory, Chile.
D.K. acknowledges support by the National Research Foundation of Korea to the Fostering Core Leaders of the Future Basic Science Program, No. 2017-002533.
M.H. acknowledges the support from Global Ph.D Fellowship Program through the National Research Foundation of Korea (NRF) funded by the Ministry of Education (NRF-2013H1A2A1033110).

{\it Facilities:} \facility{Magellan:Baade (FIRE), Gemini:South (GMOS-S)}


\begin{thebibliography}{}


\bibitem[Aihara et al.(2017a)]{Aihara17a} Aihara, H., 
Arimoto, N., Armstrong, R., et al.\ 2017, arXiv:1704.05858 

\bibitem[Aihara et al.(2017b)]{Aihara17b} Aihara, H., 
Armstrong, R., Bickerton, S., et al.\ 2017, arXiv:1702.08449 

\bibitem[Alvarez et al.(2009)]{Alvarez09} Alvarez, M.~A., Wise, J.~H., \& Abel, T.\ 2009, \apjl, 701, L133

\bibitem[Alexander \& Hickox(2012)]{Alexander12} Alexander, D.~M., 
\& Hickox, R.~C.\ 2012, New A Rev., 56, 93 

\bibitem[Ba{\~n}ados et al.(2014)]{Banados14} Ba{\~n}ados, E., 
Venemans, B.~P., Morganson, E., et al.\ 2014, \aj, 148, 14 

\bibitem[Ba{\~n}ados et al.(2016)]{Banados16} Ba{\~n}ados, E., 
Venemans, B.~P., Decarli, R., et al.\ 2016, \apjs, 227, 11 

\bibitem[Ba{\~n}ados et al.(2017)]{Banados17} Ba{\~n}ados, E., Venemans, B.~P., Mazzucchelli, C., et al.\ 2017, arXiv:1712.01860 



\bibitem[Begelman et al.(2006)]{Begelman06} Begelman, M. C., Volonteri, M., \& Rees, M. J. 2006, \mnras, 370, 289


\bibitem[Boyle et al.(2000)]{Boyle00} Boyle, B. J., Shanks, T., Croom, S. M., et al.\ 2000, \mnras,317, 1014

\bibitem[Bromm \& Loeb(2003)]{Bromm03} Bromm, V., \& Loeb, A. 2003, \apjl, 596, L34

\bibitem[Cardelli et al.(1989)]{Cardelli89} Cardelli, J.~A.,
Clayton, G.~C., \& Mathis, J.~S.\ 1989, \apj, 345, 245 


\bibitem[Coatman et al.(2016)]{Coatman16} Coatman, L., Hewett, P.~C.,
Banerji, M., \& Richards, G.~T.\ 2016, \mnras, 461, 647 

\bibitem[Coatman et al.(2017)]{Coatman17} Coatman, L., Hewett, P.~C., 
Banerji, M., et al.\ 2017, \mnras, 465, 2120 

\bibitem[Decarli et al.(2010)]{Decarli10} Decarli, R.,
 Falomo, R., Treves, A., et al.\ 2010, \mnras, 402, 2453 

\bibitem[DeGraf et al.(2017)]{DeGraf17} De Graf, C., 
Dekel, A., Gabor, J., \& Bournaud, F.\ 2017, \mnras, 466, 1462 

\bibitem[DeGraf et al.(2012)]{DeGraf12} De Graf, C., 
Di Matteo, T., Khandai, N., et al.\ 2012, \mnras, 424, 1892 


\bibitem[Denney et al.(2013)]{Denney13} Denney, K.~D., 
Pogge, R.~W., Assef, R.~J., et al.\ 2013, \apj, 775, 60 

\bibitem[De Rosa et al.(2011)]{DeRosa11} De Rosa, 
G., Decarli, R., Walter, F., et al.\ 2011, \apj, 739, 56 

\bibitem[De Rosa et al.(2014)]{DeRosa14} De Rosa, 
G., Venemans, B.~P., Decarli, R., et al.\ 2014, \apj, 790, 145 

\bibitem[Dietrich et al.(2003)]{Dietrich03} Dietrich, M.,
Hamann, F., Appenzeller, I., \& Vestergaard, M.\ 2003, \apj, 596, 817 

\bibitem[Di Matteo et al.(2012)]{DiMatteo12} Di Matteo,
 T., Khandai, N., DeGraf, C., et al.\ 2012, \apjl, 745, L29 

\bibitem[Fan et al.(2006)]{Fan06} Fan, X., Strauss, M.~A., 
Becker, R.~H., et al.\ 2006, \aj, 132, 117 

\bibitem[Fan et al.(2000)]{Fan00} Fan, X., White, R. L., 
Davis, M., et al.\ 2000, \aj, 120, 1167

\bibitem[Ferrara et al.(2014)]{Ferrara14} 
Ferrara, A., Salvadori, S., Yue, B., \& Schleicher, D.\ 2014, \mnras, 443, 2410 

\bibitem[Flesch(2015)]{Flesch15} Flesch, E.~W.\ 2015, PASA, 32, 10


\bibitem[Gallerani et al.(2017)]{Gallerani17} Gallerani, S., 
Fan, X., Maiolino, R., \& Pacucci, F.\ 2017, arXiv:1702.06123 

\bibitem[Giallongo et 
al.(2015)]{Giallongo15} Giallongo, E., Grazian, A., Fiore, F., et al.\ 2015, \aap, 578, A83 


\bibitem[Goto(2006)]{Goto06} Goto, T.\ 2006, \mnras, 371, 769 

\bibitem[Grandi(1982)]{Grandi82} Grandi, S.~A.\ 1982, \apj, 255, 25 


\bibitem[Hewett et al.(1995)]{Hewett95} Hewett, P. C., Foltz, C. B., \& Chaffee, F. H. 1995, \aj, 109, 1498

\bibitem[Ho et al.(2012)]{Ho12} Ho, L.~C., 
Goldoni, P., Dong, X.-B., Greene, J.~E., \& Ponti, G.\ 2012, \apj, 754, 11 

\bibitem[Im et al.(2007)]{Im07} Im, M., Lee, I., CHo, Y., et al. 2007, \apj, 664, 64

\bibitem[Im et al.(1997)]{Im97} Im, M., Griffiths, R.~E., 
\& Ratnatunga, K.~U.\ 1997, \apj, 475, 457 

\bibitem[Jiang et al.(2009)]{Jiang09} Jiang, L., Fan, X., Bian, 
F., et al.\ 2009, \aj, 138, 305 

\bibitem[Jiang et al.(2010)]{Jiang10} Jiang, L., Fan, X., 
Brandt, W.~N., et al.\ 2010, \nat, 464, 380 

\bibitem[Jiang et al.(2007)]{Jiang07} Jiang, L., Fan, X., 
Vestergaard, M., et al.\ 2007, \aj, 134, 1150 

\bibitem[Jiang et al.(2016)]{Jiang16} Jiang, L., 
McGreer, I.~D., Fan, X., et al.\ 2016, \apj, 833, 222 


\bibitem[Johnson et al.(2013)]{Johnson13} Johnson, J.~L., 
Whalen, D.~J., Li, H., \& Holz, D.~E.\ 2013, \apj, 771, 116 

\bibitem[Jeon et al.(2017)]{Jeon17} Jeon, Y., Im, M., Kim, D., et al.\ 2017, \apjs, 231, 16 

\bibitem[Jeon et al.(2012)]{Jeon12} Jeon, M., Pawlik, A.~H., Greif, T.~H., et al.\ 2012, \apj, 754, 34

\bibitem[Jun 
\& Im(2013)]{Jun13} Jun, H.~D., \& Im, M.\ 2013, \apj, 779, 104 

\bibitem[Jun et al.(2015)]{Jun15} Jun, H.~D., Im, M., Lee, 
H.~M., et al.\ 2015, \apj, 806, 109 

\bibitem[Jun et al.(2017)]{Jun17} Jun, H.~D., Im, M., Kim, D., \& Stern, D.\ 2017, \apj, 838, 41 


\bibitem[Karouzos et al.(2015)]{Karouzos15} Karouzos, M., 
Woo, J.-H., Matsuoka, K., et al.\ 2015, \apj, 815, 128 

\bibitem[Kashikawa et al.(2015)]{Kashikawa15} Kashikawa, N., 
Ishizaki, Y., Willott, C.~J., et al.\ 2015, \apj, 798, 28 



\bibitem[Kim et al.(2010)]{Kim10} Kim, D., Im, M., \& Kim, M.\ 2010, \apj, 724, 386 


\bibitem[Kim et al.(2015)]{Kim15} Kim, Y., Im, M.,
Jeon, Y., et al.\ 2015, \apjl, 813, L35 


\bibitem[Kurk et al.(2009)]{Kurk09} Kurk, J.~D., Walter, F., Fan, X., et al.\ 2009, \apj, 702, 833 

\bibitem[Kurk et al.(2007)]{Kurk07} Kurk, J.~D., Walter, F., Fan, X., et al.\ 2007, \apj, 669, 32 

\bibitem[Lee et al.(2008)]{Lee08} Lee, I., Im, M., Kim, M., et al. 2008, \apjs, 175, 116

\bibitem[Li(2012)]{Li12} Li, L.-X. 2012, \mnras, 424, 1461


\bibitem[Lodato \& Natarajan(2006)]{Lodato06} Lodato, G., \& Natarajan, P. 2006, \mnras, 371, 1813

\bibitem[Lusso et al.(2012)]{Lusso12} Lusso, E., Comastri, A., Simmons, B.~D., et al.\ 2012, \mnras, 425, 623 


\bibitem[Madau et al.(2014)]{Madau14} Madau, P., Haardt, F., \& Dotti, M.\ 2014, \apjl, 784, L38

\bibitem[Markwardt(2009)]{Markwardt09} Markwardt, C.~B.\ 2009,
Astronomical Data Analysis Software and Systems XVIII, 411, 251

\bibitem[Matsuoka et al.(2016)]{Matsuoka16} Matsuoka, Y., 
Onoue, M., Kashikawa, N., et al.\ 2016, \apj, 828, 26 

\bibitem[Matsuoka et al.(2017)]{Matsuoka17} Matsuoka, Y., 
Onoue, M., Kashikawa, N., et al.\ 2017, arXiv:1704.05854 

\bibitem[Mazzucchelli et al.(2017)]{Mazzucchelli17} Mazzucchelli, 
C., Ba{\~n}ados, E., Venemans, B.~P., et al.\ 2017, arXiv:1710.01251 

\bibitem[Milosavljevi{\'c} et al.(2009)]{Milosavljevic09} Milosavljevi{\'c}, M., Couch, S.~M., \& Bromm, V.\ 2009, \apjl, 696, L146 

\bibitem[Mineshige et al.(2000)]{Mineshige00} Mineshige, S., Kawaguchi, T., Takeuchi, M., \& Hayashida, K.\ 2000, \pasj, 52, 499 

\bibitem[Mortlock et al.(2011)]{Mortlock11} Mortlock, D.~J., 
Warren, S.~J., Venemans, B.~P., et al.\ 2011, \nat, 474, 616 

\bibitem[Ohsuga et al.(2005)]{Ohsuga05} Ohsuga, K., Mori, M., Nakamoto, T., \& Mineshige, S.\ 2005, \apj, 628, 368 


\bibitem[Pacucci et al.(2015)]{Pacucci15} 
 Pacucci, F., Volonteri, M., \& Ferrara, A.\ 2015, \mnras, 452, 1922 
 
\bibitem[P{\^a}ris et al.(2017)]{Paris17} P{\^a}ris, I.,
 Petitjean, P., Ross, N.~P., et al.\ 2017, \aap, 597, A79 

\bibitem[Park et al.(2013)]{Park13} Park, D., 
Woo, J.-H., Denney, K.~D., \& Shin, J.\ 2013, \apj, 770, 87 

\bibitem[Park et al.(2017)]{Park17} Park, D., 
Barth, A.~J., Woo, J.-H., et al.\ 2017, \apj, 839, 93 

\bibitem[Park \& Ricotti(2012)]{Park12} Park, K., \& Ricotti, M.\ 2012, \apj, 747, 9

\bibitem[Pelupessy et al.(2007)]{Pelupessy07} Pelupessy, F.~I., Di Matteo, T., \& Ciardi, B.\ 2007, \apj, 665, 107 


\bibitem[Pezzulli et al.(2016)]{Pezzulli16} 
Pezzulli, E., Valiante, R., \& Schneider, R.\ 2016, \mnras, 458, 3047

\bibitem[Regan et al.(2017)]{Regan17} Regan, J.~A., 
Visbal, E., Wise, J.~H., et al.\ 2017, Nature Astronomy, 1, 0075 

\bibitem[Richards et al.(2006)]{Richards06} Richards, G.~T., 
Strauss, M.~A., Fan, X., et al.\ 2006, \aj, 131, 2766 


\bibitem[Runnoe et al.(2012)]{Runnoe12} Runnoe, J.~C., 
Brotherton, M.~S., \& Shang, Z.\ 2012, \mnras, 422, 478 

\bibitem[Runnoe et al.(2013)]{Runnoe13} Runnoe, J.~C., Brotherton, M.~S., 
Shang, Z., \& DiPompeo, M.~A.\ 2013, \mnras, 434, 848 

\bibitem[Sakurai et al. (2016)]{Sakurai16} 
Sakurai, Y., Inayoshi, K., \& Haiman, Z., \mnras, 461, 4496

\bibitem[Schlafly \& Finkbeiner(2011)]{Schlafly11} Schlafly, E.~F., 
\& Finkbeiner, D.~P.\ 2011, \apj, 737, 103 


\bibitem[Schmidt \& Green(1983)]{Schmidt83} Schmidt, M., \& Green, R. F. 1983, \apj, 269, 352

\bibitem[Schulze et al.(2015)]{Schulze15} Schulze, A., 
Bongiorno, A., Gavignaud, I., et al.\ 2015, \mnras, 447, 2085 

\bibitem[Shen et al.(2008)]{Shen08} Shen, Y., Greene, J.~E., 
Strauss, M.~A., Richards, G.~T., \& Schneider, D.~P.\ 2008, \apj, 680, 169-190 

\bibitem[Shen et al.(2011)]{Shen11} Shen, Y., 
Richards, G.~T., Strauss, M.~A., et al.\ 2011, \apjs, 194, 45 


\bibitem[Smidt et al.(2017)]{Smidt17} Smidt, J., Whalen, D.~J., 
Johnson, J.~L., \& Li, H.\ 2017, arXiv:1703.00449 

\bibitem[Smole et al.(2015)]{Smole15} Smole, M., 
Micic, M., \& Martinovi{\'c}, N.\ 2015, \mnras, 451, 1964 

\bibitem[Tang et al.(2012)]{Tang12} Tang, B., 
Shang, Z., Gu, Q., Brotherton, M.~S., \& Runnoe, J.~C.\ 2012, \apjs, 201, 38 

\bibitem[Tody(1993)]{Tody93} Tody, D.\ 1993, Astronomical Data Analysis Software and Systems II, 52, 173 

\bibitem[Trakhtenbrot \& Netzer(2012)]{Trakhtenbrot12} Trakhtenbrot, B., 
\& Netzer, H.\ 2012, \mnras, 427, 3081 

\bibitem[Venemans et al.(2013)]{Venemans13} Venemans, B.~P., 
Findlay, J.~R., Sutherland, W.~J., et al.\ 2013, \apj, 779, 24 

\bibitem[Venemans et al.(2015a)]{Venemans15a} Venemans, B.~P., 
Ba{\~n}ados, E., Decarli, R., et al.\ 2015a, \apjl, 801, L11 

\bibitem[Venemans et al.(2015b)]{Venemans15b} Venemans, B.~P., 
Verdoes Kleijn, G.~A., Mwebaze, J., et al.\ 2015b, \mnras, 453, 2259 

\bibitem[Venemans et al.(2016)]{Venemans16} Venemans, B.~P., 
Walter, F., Zschaechner, L., et al.\ 2016, \apj, 816, 37 

\bibitem[Vestergaard 
\& Osmer(2009)]{Vestergaard09} Vestergaard, M., \& Osmer, P.~S.\ 2009, \apj, 699, 800

\bibitem[Vestergaard \& Peterson(2006)]{Vestergaard06} Vestergaard, M., 
\& Peterson, B.~M.\ 2006, \apj, 641, 689 

\bibitem[Volonteri(2012)]{Volonteri12} Volonteri, M.\ 2012, Science, 337, 544 

\bibitem[Volonteri et al.(2008)]{Volonteri08} Volonteri, M., 
Lodato, G., \& Natarajan, P.\ 2008, \mnras, 383, 1079 

\bibitem[Volonteri \& Rees(2005)]{Volonteri05} 
Volonteri, M., \& Rees, M. J.\ 2005, \apj, 633, 624

\bibitem[Volonteri et al.(2015)]{Volonteri15} Volonteri, M., Silk, J., \& Dubus, G.\ 2015, \apj, 804, 148 

\bibitem[Watarai et al.(2001)]{Watarai01} Watarai, K.-y., Mizuno, T., \& Mineshige, S.\ 2001, \apjl, 549, L77 

\bibitem[Wang \& Netzer(2003)]{Wang03} Wang, J.-M., \& Netzer, H.\ 2003, \aap, 398, 927 

\bibitem[Willott et al.(2003)]{Willott03} Willott, C.~J., 
McLure, R.~J., \& Jarvis, M.~J.\ 2003, \apjl, 587, L15 

\bibitem[Willott et al.(2010a)]{Willott10a} Willott, C.~J., Albert, 
L., Arzoumanian, D., et al.\ 2010a, \aj, 140, 546 


\bibitem[Willott et al.(2010b)]{Willott10b} Willott, C.~J., 
Delorme, P., Reyl{\'e}, C., et al.\ 2010b, \aj, 139, 906 


\bibitem[Wyithe \& Loeb(2012)]{Wyithe12} Wyithe, J. S. B., \& Loeb, A. 2012, \mnras, 425, 2892

\bibitem[Woo et al.(2013)]{Woo13} Woo, J.-H., Schulze, A., Park, D., et al.\ 2013, \apj, 772, 49 

\bibitem[Wu et al.(2015)]{Wu15} Wu, X.-B., Wang, F., Fan, 
X., et al.\ 2015, \nat, 518, 512 


\end{thebibliography}
\end{document}